\documentclass[referee]{raa}            

\usepackage{graphicx}
\usepackage{float}
\usepackage{subfig}
\usepackage{lineno}
\begin{document}

   \title{An algorithm to resolve $\gamma$-rays from charged cosmic rays with DAMPE
}

   \volnopage{Vol.0 (200x) No.0, 000--000}      
   \setcounter{page}{1}          

   \author{Z. L. Xu\inst{1,2}
    \and K. K.  Duan\inst{1,2}
    \and Z. Q. Shen\inst{1,2}
    \and S. J. Lei\inst{1}
    \and T. K. Dong\inst{1}
    \and F. Gargano\inst{3}
    \and S. Garrappa\inst{4,5}
    \and D. Y. Guo\inst{6}
    \and W. Jiang\inst{1,2}
    \and X. Li\inst{1} \footnote{Corresponding author:xiangli@pmo.ac.cn}
    \and Y. F. Liang\inst{1} \footnote{Corresponding author:liangyf@pmo.ac.cn}
    \and M. N. Mazziotta\inst{3}
    \and M. M. Salinas\inst{7}
    \and M. Su\inst{1}
    \and V. Vagelli\inst{4,5}
    \and Q. Yuan\inst{1,8}
    \and C. Yue\inst{1,2}
    \and J. J. Zang\inst{1} \footnote{Corresponding author:zangjj@pmo.ac.cn}
    \and Y. P. Zhang\inst{9}
    \and Y. L. Zhang\inst{10}
    \and S. Zimmer\inst{7}
   }

   \institute{ Key Laboratory of Dark Matter and Space Astronomy, Purple Mountain Observatory, Chinese Academy of Sciences, 2 West Beijing Road, Nanjing 210008, China; {\it xuzl@pmo.ac.cn}\\
\and
University of Chinese Academy of Sciences, 19 Yuquan Road, Beijing 100049, China\\
\and
Istituto Nazionale di Fisica Nucleare Sezione di Bari, I-70125, Bari, Italy\\
\and
Istituto Nazionale di Fisica Nucleare Sezione di Perugia, I-06123 Perugia, Italy\\
\and
Dipartimento di Fisica e Geologia, Universit\`a degli Studi di Perugia, I-06123 Perugia, Italy\\
\and
Institute of High Energy Physics, Chinese Academy of Sciences, YuquanLu 19B, Beijing 100049, China\\
\and
Department of Nuclear and Particle Physics, University of Geneva, CH-1211, Switzerland\\
\and
School of Astronomy and Space Science, University of Science and Technology of China, Hefei 230026, China\\
\and
Institute of Modern Physics, Chinese Academy of Sciences, 509 Nanchang Road, Lanzhou 730000, China\\
\and
State Key Laboratory of Particle Detection and Electronics, University of Science and Technology of China, Hefei 230026, China\\
   }

   \date{Received~~2009 month day; accepted~~2009~~month day}

\abstract{ The DArk Matter Particle Explorer (DAMPE), also known as Wukong in China, launched on December 17, 2015, is a new high energy cosmic ray and $\gamma$-ray satellite-borne observatory in space. One of the main scientific goals of DAMPE is to observe GeV-TeV high energy $\gamma$-rays with accurate energy, angular, and time resolution, to indirectly search for dark matter particles and for the study of high energy astrophysics. Due to the comparatively higher fluxes of charged cosmic rays with respect to $\gamma$-rays, it is challenging to identify $\gamma$-rays with sufficiently high efficiency minimizing the amount of charged cosmic ray contamination. In this work we present a method to identify $\gamma$-rays in DAMPE data based on Monte Carlo simulations, using the powerful electromagnetic/hadronic shower discrimination provided by the calorimeter and the veto detection of charged particles provided by the plastic scintillation detector. Monte Carlo simulations show that after this selection
the number of electrons and protons that contaminate the selected $\gamma$-ray events at $\sim10$ GeV amounts to less than 1\% of the selected sample. Finally, we use flight data to verify the effectiveness of the method by highlighting known $\gamma$-ray sources in the sky and by reconstructing preliminary light curves of the Geminga pulsar.
}

   \authorrunning{Z.-L. Xu et al. }            
   \titlerunning{An algorithm to resolve $\gamma$-rays from charged cosmic rays with DAMPE}  

   \maketitle

%
%
\section{Introduction}           
\label{sect:intro}
The DArk Matter Particle Explorer (DAMPE) is a satellite-borne, general-purpose high energy cosmic ray (CR) and $\gamma$-ray observatory (Chang ~\cite{Chang14}; Chang et al. ~\cite{Chang17}). It was launched on December 17, 2015, and entered normal science operations on December 27th, 2015.

\begin{figure}
\centering
\includegraphics[width=130mm,height=90mm]{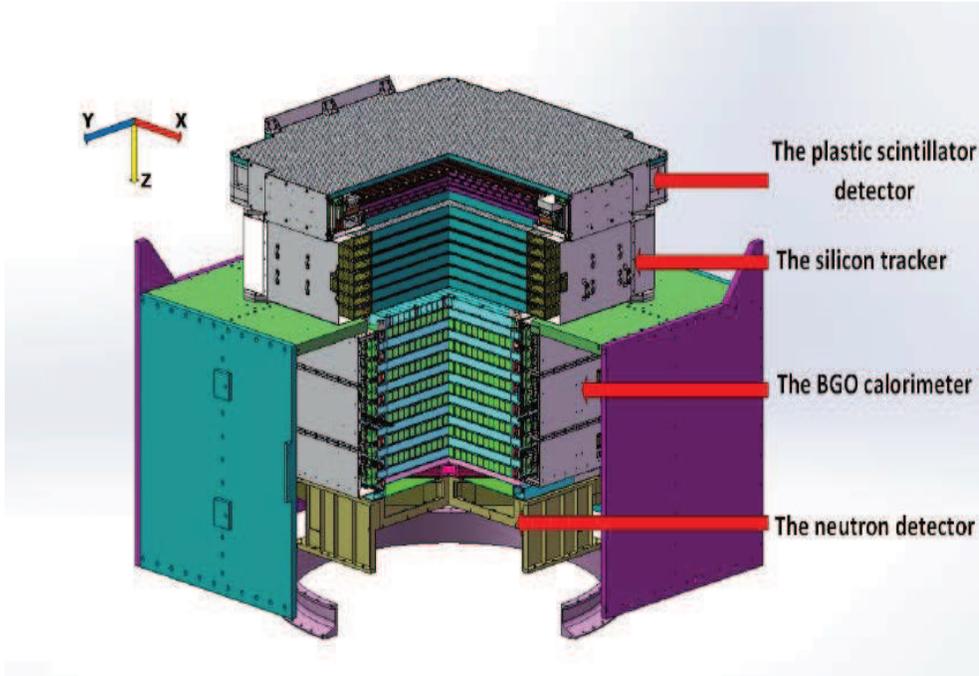}
\caption{Schematic view of the DAMPE detector.}
\label{Fig1}
\end{figure}

Fig.~\ref{Fig1} shows a schematic view of the DAMPE detector. From top to bottom, DAMPE is composed of four sub-detectors: a Plastic Scintillation Detector (PSD), a Silicon-Tungsten tracKer converter detector (STK), a BGO calorimeter (BGO), and a Neutron Detector (NUD)(Chang ~\cite{Chang14}; Chang et al. ~\cite{Chang17}). The PSD has an active area of $\rm 82.5 \times 82.5\, cm^2$ and features two layers of plastic scintillating bars with 2.8 cm width, arranged along $X$ and $Y$ directions, respectively. It acts as a charge detector for charged CRs and as an Anti-Coincidence Detector (ACD) for $\gamma$-rays. The STK is made of six tracking planes, each consisting of two sub-layers of single side silicon strip detectors orthogonally arranged. The primary goal of STK is to determine the direction of incident particles. Three tungsten foils, each 1 mm thick, are interleaved with the top three planes in order to increase the pair-conversion efficiency for a total radiation length in the STK of 0.976 X0. The BGO calorimeter is composed of 308 BGO scintillation crystals with $2.5\times2.5$\,cm$^{2}$ square cross section, arranged hodoscopically in 14 layers, to measure the three-dimensional shower profile of particles. The BGO calorimeter covers an area of $\rm 60 \times 60\,cm^2$ for a total depth of 32 X0 for full containment of TeV electromagnetic showers. In addition to the measurement of the particle energy, the analysis of the BGO image is used to differentiate between electromagnetic and hadronic showers and to provide an additional measurement of the particle incoming direction. Finally, the NUD detects delayed neutrons that arise from a proton induced particle cascade to consequently improve the electron/proton (e/p) separation at TeV energies.

The intensity of the flux of charged CRs at Earth dominates over that of $\gamma$-rays by several orders of magnitude (Barwick et al. ~\cite{Barwick98}). Therefore, a large amount of charged particles must be efficiently suppressed to minimize their contamination in the $\gamma$-ray sample while retaining a high $\gamma$-ray selection efficiency.
In this work we develop a possible procedure to select $\gamma$-rays in the DAMPE data. The basic idea is to combine the electromagnetic/hadronic shower discrimination provided by the BGO calorimeter with the capability of the PSD to measure the charge of incoming particles and thus act as veto. We show that this method can reach a high selection efficiency while keeping a low background contamination.

This paper is organized as follows. In Sec. 2 we describe the detailed $\gamma$-ray selection procedure, including the e/p separation using BGO (Sec.~2.1), the STK track selection (Sec.~2.2), and the charged particle
rejection using PSD (Sec.~2.3).
In Sec. 3 we discuss the acceptance of $\gamma$-rays and residual electrons after the selection algorithm.
In Sec.~4, we test the performances of the $\gamma$-ray selection on flight data by highlighting known $\gamma$-ray sources and by reconstructing of the phase diagram of the Geminga Pulsar as a testbench. Finally, we summarize our work in Sec.~5.

\section{Resolving $\gamma$-rays from charged particles}
$\gamma$-rays crossing the fiducial volume of the detector can be classified according to their interactions with the detector materials above the BGO calorimeter. Most of the $\gamma$-rays convert in highly collimated $e^+e^-$ pairs in the material above the BGO -- especially in the tungsten layers of the STK -- resulting in only one track reconstructed in the STK, while a smaller fraction of $\gamma$-rays reach the BGO without converting before. In this work, we focus on the identification of the first, more abundant, class of $\gamma$-rays. The identification of non-converting $\gamma$-rays will not be covered in this contribution, and will be the subject of a future publication.

The signature of a converting $\gamma$-ray crossing the DAMPE instrument is an electromagnetic shower in the BGO matched by a track in lower part of the STK together with the absence of any relevant signal released in the surrounding area of the PSD. However, the finite detection efficiency of the PSD coupled with systematic effects on the track reconstruction or particle identification may lead to a residual amount of charged particles that may be misidentified as $\gamma$-rays. To minimize this source of contamination we first use the BGO to reject hadronic CRs crossing the BGO fiducial volume and we subsequently use the PSD as ACD for residual charged CRs. The $\gamma$-ray selection procedure developed here proceeds as follows:

i) identification of the shower in the BGO, reject protons using its topological development;

ii) search for the conversion track in the STK matching with the BGO shower axis;

iii) search for the PSD cluster associated to the primary CRs track and veto of events with relevant energy deposit in the cluster to reject electrons and residual hadrons in the selected BGO samples;

The procedure for $\gamma$-ray identification has been defined and studied using Monte Carlo (MC) simulation data as described in details later. The selection requirements have been defined to be loose to maximize the effective $\gamma$-ray acceptance. We only require the selected STK track to pass through the PSD and through the top four layers of BGO, instead of penetrating all the 14 BGO layers. We will show later that such approach limits the amount of charged particle background below an optimal threshold while increasing the acceptance for $\gamma$-ray.
\subsection{Electron/proton separation}
The natural flux of protons is already suppressed at trigger level by the so-called ``hardware suppression''. The differences in the total energy deposit in the BGO between hadrons and electrons/$\gamma$-rays result in different trigger efficiencies. As a consequence of this, the natural flux of protons is suppressed by a factor of ~20 in the triggered data (Chang ~\cite{Chang14}, Chang et al. ~\cite{Chang17}). The majority of protons and nuclei in the DAMPE data are instead removed from the $\gamma$-ray samples using the analysis of the topological development of the shower in the BGO calorimeter.

We use extensive MC simulations of the entire DAMPE detector (including support structures) using the GEANT 4.10.02 version (Agostinelli et al. ~\cite{Agostinelli03}; Wang et al. ~\cite{Wang17}) to define the e/p separation cuts\footnote{The hadronic model QGSP\_FTFP\_BERT is used to generate proton sample used in this analysis.}. We perform a topological selection on the BGO energy deposit to identify electromagnetic showers analyzing both the longitudinal ( (a) and (b) ) and lateral ( (c) and (d) ) development of the shower.

(a) The number of signal crystals in the top four layers of the BGO ($N_l$, $l$ is the layer number) are required to be larger than two since protons usually shower later than electrons.

\begin{figure}
\centering
\includegraphics[width=130mm,height=90mm]{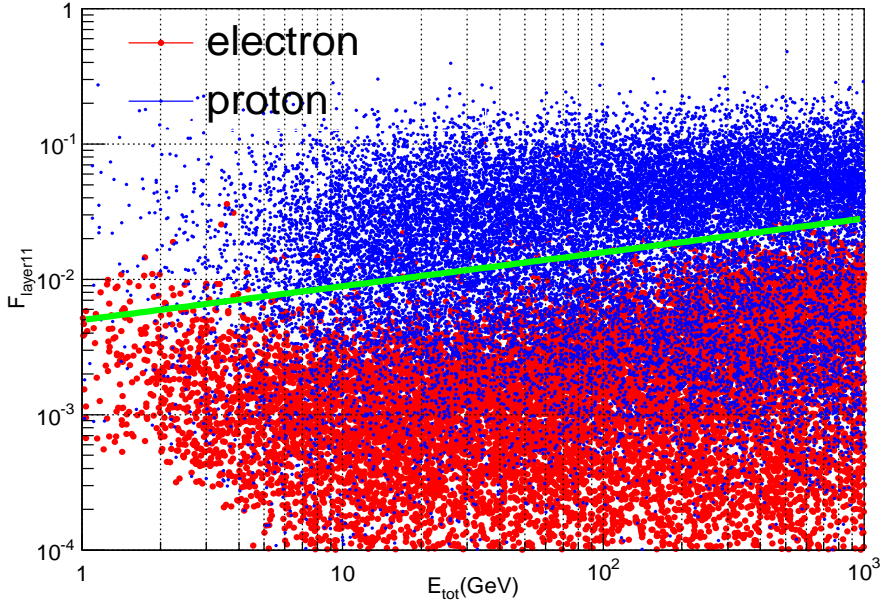}
\caption{Energy deposit ratio for layer 11 as function of total deposited energy in the BGO, for electrons (red) and protons (blue). The green line indicates the cut to separate electrons and protons.}
\label{Fig2}
\end{figure}

(b) Since electrons develop earlier than protons, and their energy deposit fractions in the bottom layers of BGO are significantly smaller than those of protons, we use the energy deposit ratio in each layer $\rm F_l$ (the energy deposited in layer $\rm l$ divided by the total deposited energy in the BGO, $\rm E_{tot}$) as discriminating variables. First, in order to remove events crossing DAMPE from the BGO sides we require $\rm F_l<0.45$ for all layers. We further set an energy- and layer-dependent upper threshold to the values of $\rm F_l$ in the bottom four BGO layers.
\begin{eqnarray}
F_l \;<\; 0.005 \times {\rm (E_{tot}/GeV)}^{(0.8-0.05\times \rm l)}
\end{eqnarray}
where $\rm E_{tot}$ is the total energy deposit in the BGO in GeV.
Fig.~\ref{Fig2} shows the energy dependence of $\rm F_{layer11}$ for electrons and protons, as well as the selection cut.

\begin{figure}
\centering
\includegraphics[width=130mm,height=90mm]{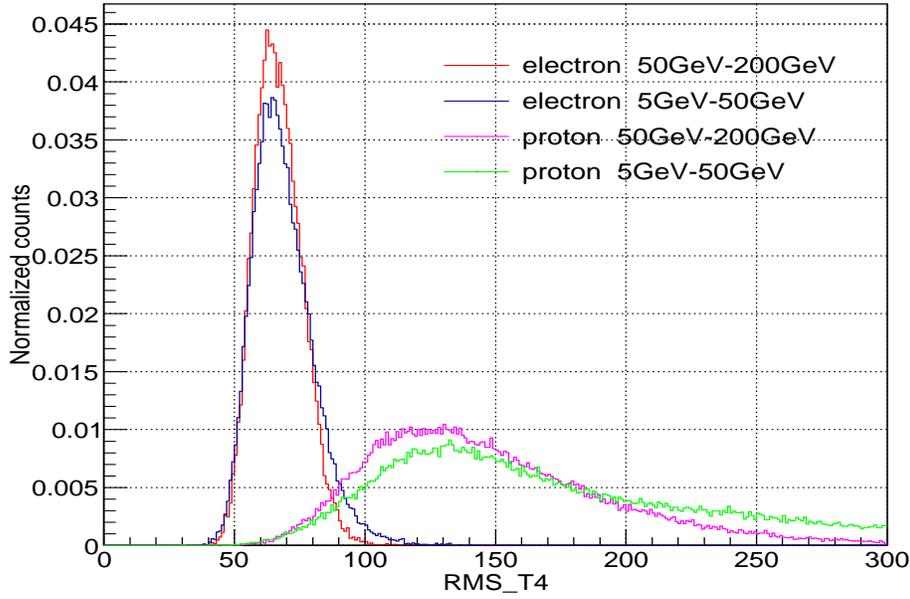}
\caption{Distributions of RMS\_T4 for electrons and protons in the deposited energy ranges 5-50 GeV and 50-200 GeV.}
\label{Fig3}
\end{figure}

(c) Since the transverse spread of the electromagnetic shower in the BGO for electrons is much smaller than that of protons, the energy-weighted root-mean-square (RMS) value of hit
positions in each layer $\mathrm{RMS}_{l}$ represents an estabilished discriminant variable (Chang et al. ~\cite{Chang08}) based on the shower lateral development to discriminate between electromagnetic and hadronic showers. Since we only select particles traversing at least four BGO layers, we use in this work the sum of the RMS of the first four layers (``RMS\_T4'') as discriminant. As shown in Fig.~\ref{Fig3}, the RMS\_T4 distribution of electrons assume values significantly smaller than that of protons and both distributions do not strongly depend on the deposited energy. We select events for which $\mathrm{RMS\_T4}<100$. This selection yields in fact a good e/p separation almost independent from the energy of the incoming particles.

\begin{figure}
\centering
\includegraphics[width=130mm,height=90mm]{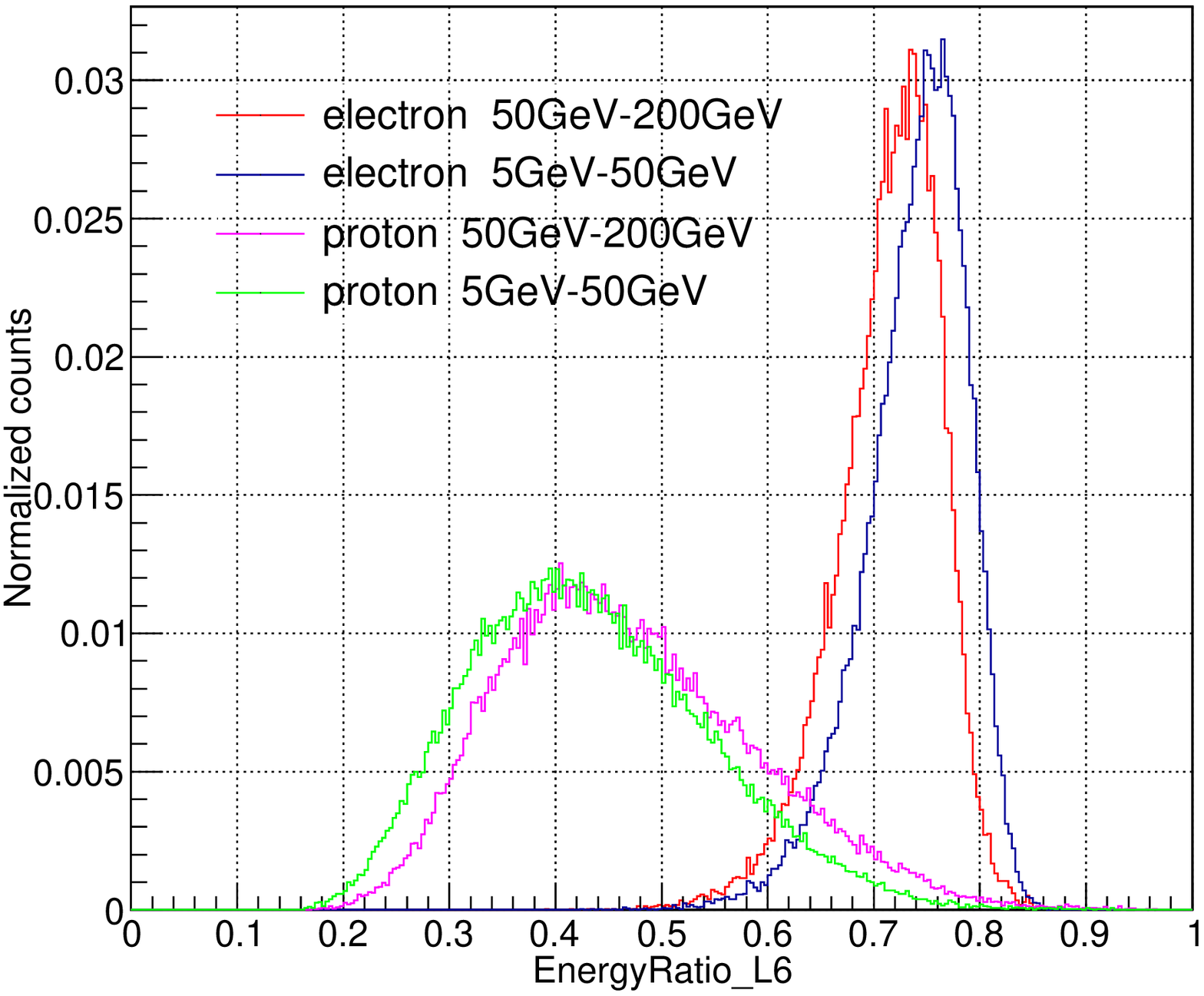}
\caption{Distributions of EnergyRatio\_L6 for electrons and protons in the deposited energy ranges 5-50 GeV and 50-200GeV.}
\label{Fig4}
\end{figure}

(d) The energy deposit of electromagnetic showers is concentrated for 90\% in one Moliere radius (corresponding to about 2.3~cm in BGO crystals). In this work, we further characterize the lateral profile of the energy distribution using the energy ratio of the six bars with the largest energy deposit to the total energy deposit in the BGO (``EnergyRatio\_L6''). Fig.~\ref{Fig4} shows the distributions of such variable for electrons and protons. We apply lower threshold ($\mathrm{EnergyRatio\_L6} > 0.5$) to remove the bulk of the proton background, with an efficiency on electron close to 100\% in the whole energy range.

\begin{figure*}[!ht]
  \centering
  \subfloat[]{
    \includegraphics[width=.5\textwidth,height=80mm]{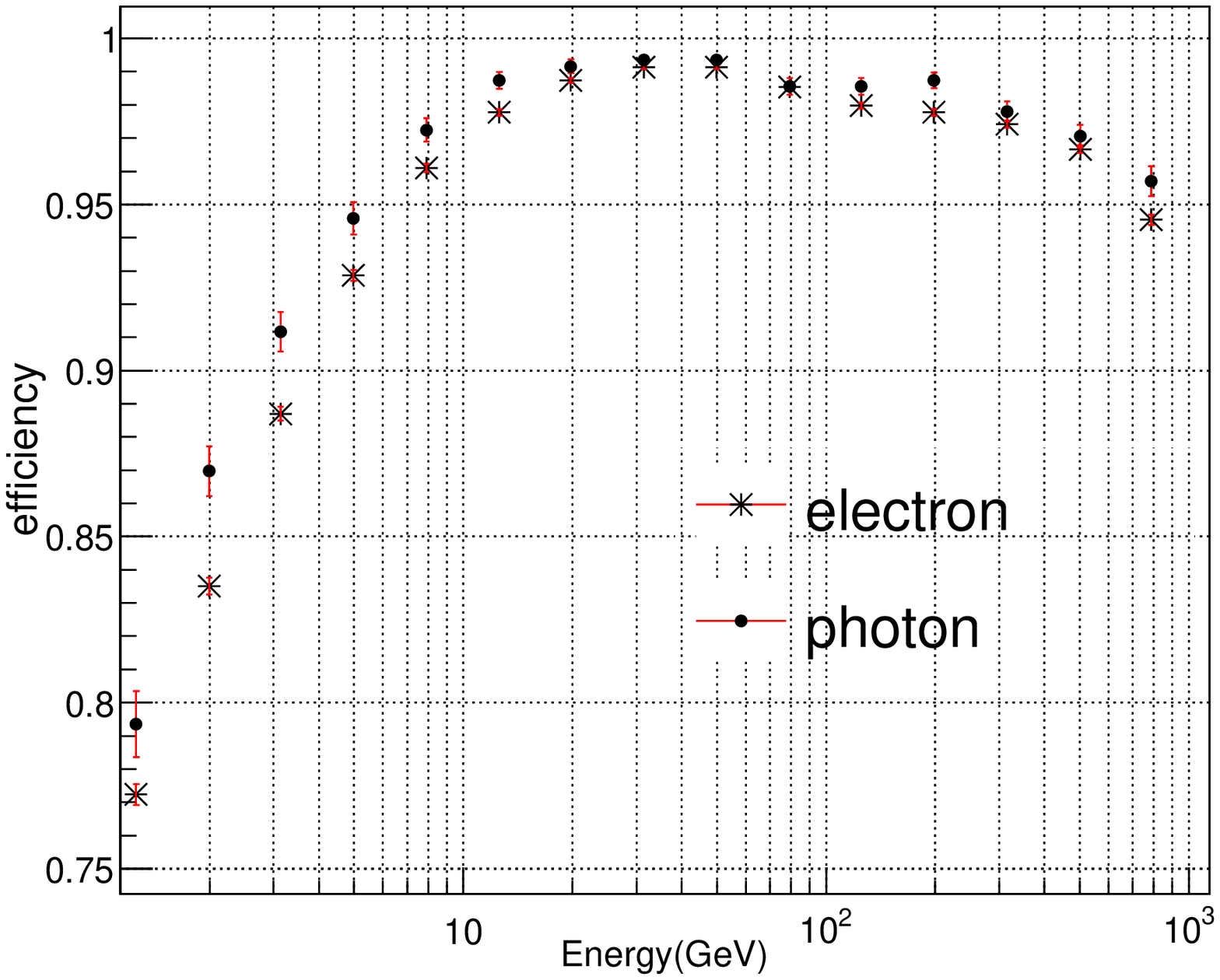}
  }
  \subfloat[]{
    \includegraphics[width=.5\textwidth,height=80mm]{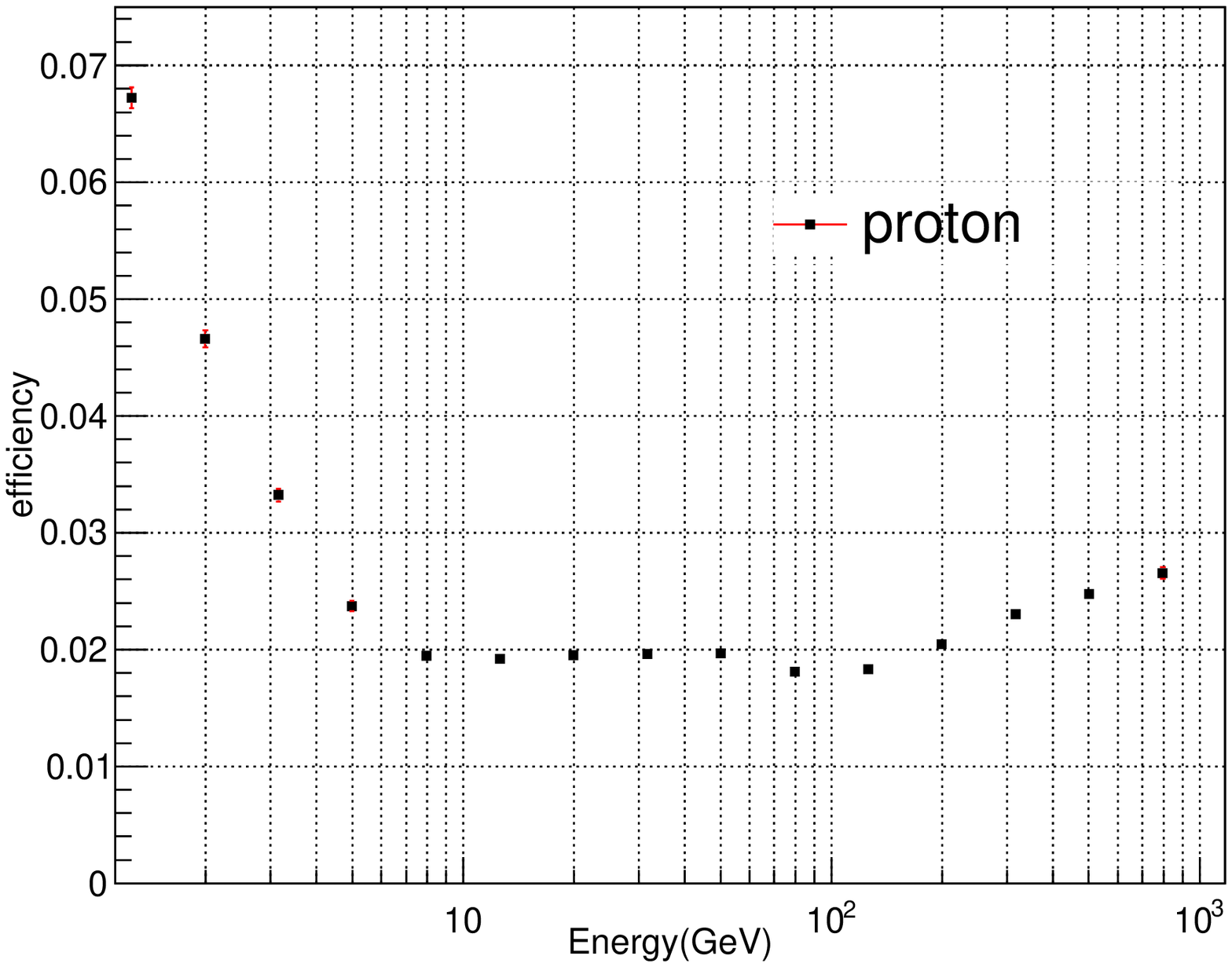}
  }
  \caption{Particle identification efficiency for electrons, $\gamma$-rays(left) and protons(right) after the electron selection with the BGO.}
  \label{Fig5}
\end{figure*}

Fig.~\ref{Fig5} shows the particle identification efficiency as functions of the deposited energy in the BGO. Only a few percents of protons are misidentified as electrons/$\gamma$-rays after the e/p BGO selection. Together with the ``hardware suppression'', a total of about $99.9\%$ of protons have been suppressed at this stage, while $\sim95\%$ electrons and $\gamma$-rays are kept for energies higher than a few GeVs. $\gamma$-rays induce $e^+e^-$ pairs tend to shower earlier in the BGO calorimeter, resulting in a slightly higher detection efficiency by few percents of $\gamma$-rays than electrons especially for the lowest energies.

\subsection{STK track selection}
Tracks in the STK are reconstructed using a modified Kalman filter method while the BGO direction is reconstructed using an energy centroid method (Chang et al. ~\cite{Chang17}).
The finite spatial resolution of the BGO track (Chang et al. ~\cite{Chang17}) may introduce systematic effects in the selection of the PSD bars to be used as veto, which may result in a relevant number of charged particles misidentified as $\gamma$-rays. In the strategy outlined in this contribution only $\gamma$-rays converting in the STK are considered. For converting $\gamma$-ray events, due to backscattering and other secondary particle interactions, typically more than one track is reconstructed in the STK and the energy deposit in the PSD is typically more abundant with respect to non showering CRs
\footnote{The intensity of this effect increases towards higher energies in which the higher track multiplicity due to backscattering events from the BGO increase the combinatorial background in each event.}.
Therefore, in order to effectively use the PSD as ACD, the correct STK track needs to be assigned to the primary incoming particle for each event.

We have developed an algorithm to select the primary STK track based on the following considerations. First, the track needs to span through several STK layers, since in general the incident primary particle is much more energetic than secondary particles (which typically yield shorter tracks). Second, the primary STK track should match the BGO shower axis, while backscattering particles could yield to secondary tracks that are much less spatially correlated with the primary track. Finally, in case that the electromagnetic shower starts in the STK, the energy deposit for electron/$\gamma$-ray events should also concentrate along the primary track. Based on the above considerations, we define an empirical variable ${\rm TQ}$ to describe the track quality
\begin{eqnarray}
{\rm TQ} = \left(\frac{1+{\rm E_{r}}}{\ln ({\rm D_{sum}}/{\rm mm})}
\right) \times \left(1+\frac{{\rm N_{tr}}-3}{12}\right).
\end{eqnarray}

Where $\rm E_{r}$ is to the ratio between the energy deposited within a 5mm cylinder around a track candidate and the total deposited energy in the STK, $\rm D_{sum}$ is the sum of the distances between the center-of-gravity in the first four BGO layers and the positions obtained extrapolating the STK track to the corresponding BGO layers, and N{tr} is the number of hits used in the STK track reconstruction. The track with the maximum value of TQ is identified as the primary track and is associated with the shower in the BGO. Events with no reconstructed tracks in the STK are not selected to be a $\gamma$-ray candidate event in this approach.

\begin{figure*}[!ht]
  \centering
  \subfloat[]{
    \includegraphics[width=.5\textwidth,height=80mm]{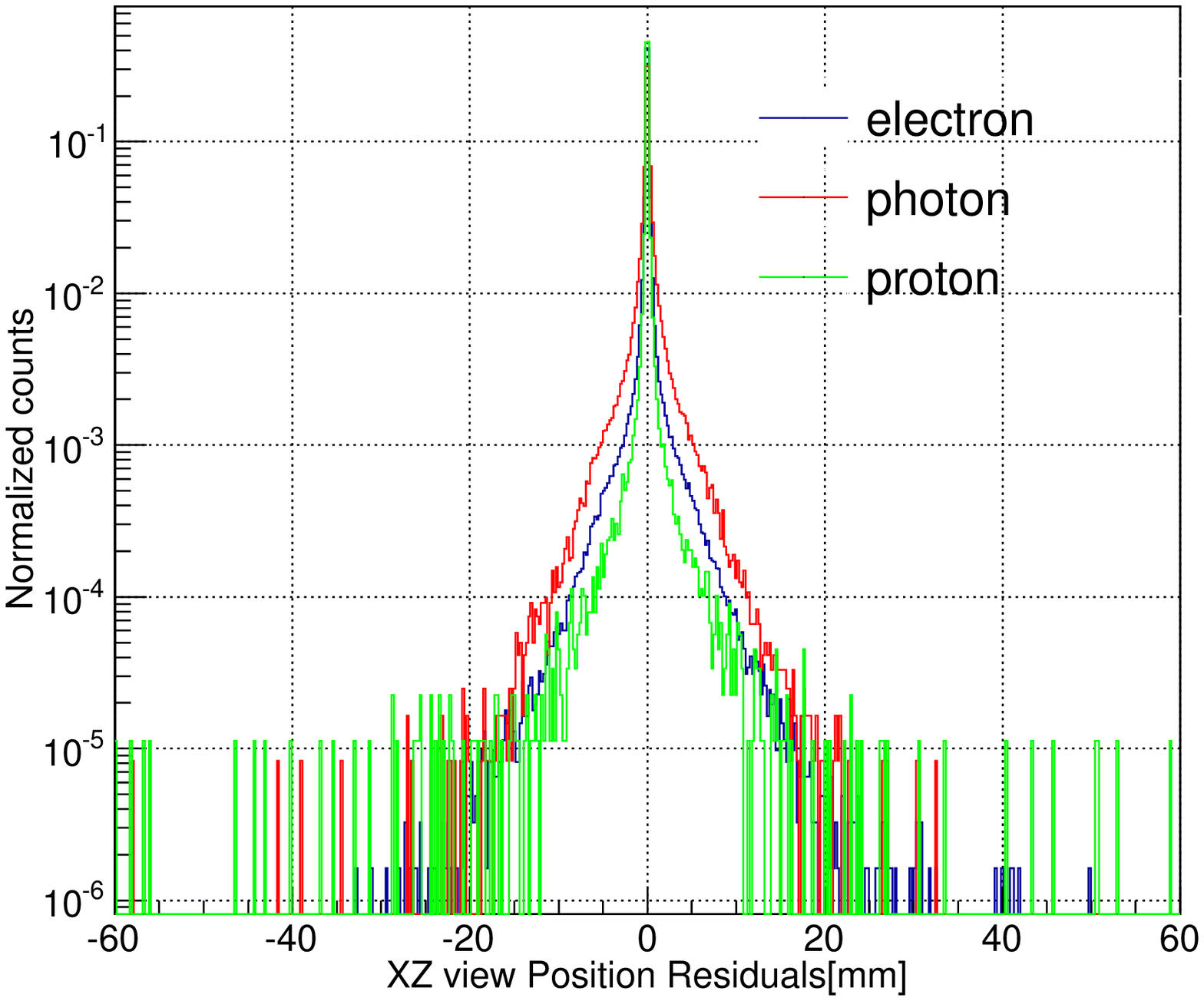}
  }
  \subfloat[]{
    \includegraphics[width=.5\textwidth,height=80mm]{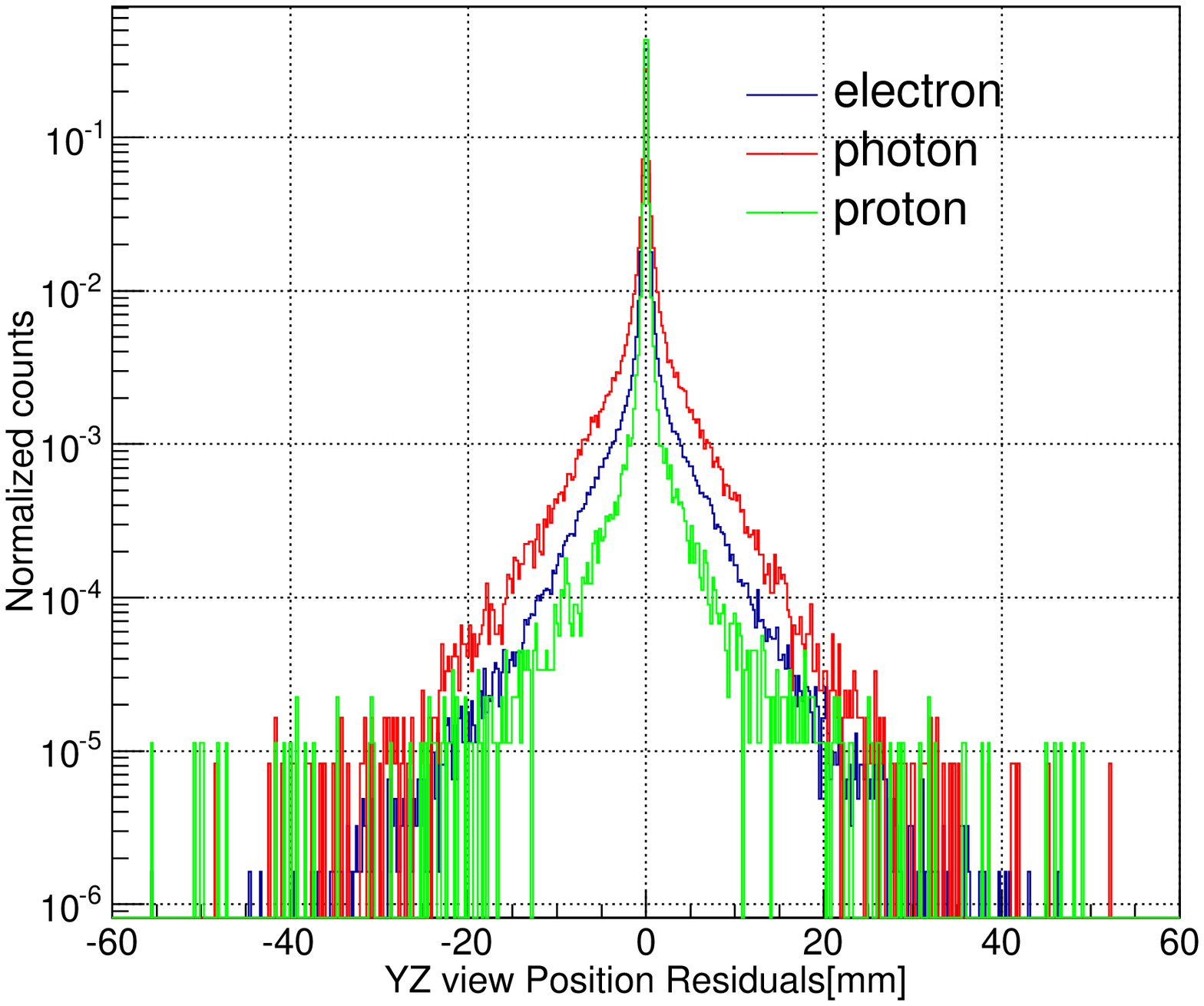}
  }
  \caption{Distribution for the STK track residuals in the $XZ$(left) and $YZ$(right) plane of the PSD, for selected STK tracks of protons (green), electrons (blue), and $\gamma$-rays (red).}
  \label{Fig6}
\end{figure*}

The primary track is then extrapolated to the PSD volume in order to verify the presence of activity in the vicinity of the primary particle crossing coordinate. We show in Fig.~\ref{Fig6} the distributions of the STK position residuals, defined as the difference between the true impact point in the PSD and that predicted by extrapolating the STK reconstructed track in the central Z coordinate of the PSD.
Less than $10^{-4}$ of events have position residuals larger than 28 mm corresponding to the lateral width of one PSD bar. Therefore relatively few PSD bars have to be checked to search for energy deposits for the charged particle veto. Since backscattered secondaries from the BGO shower may deposit energy in a wide area of the PSD, the number of PSD bars used to veto charged particles has to be minimized to maximize the efficiency for $\gamma$-ray selection. In this work, only the bar crossed by the STK track together with the two adjacent bars are used to check the veto conditions.

\subsection{Charged particle rejection with PSD}

The detection efficiency of PSD is very important to evaluate its background suppression. The efficiency is measured using minimum ionization particle (MIP) events of non-showering protons collected during flight operations. We define the detection
efficiency for charged particles of every PSD bar as
\begin{eqnarray}
\eta = \frac{N_{\rm Signal}}{N_{\rm Total}}
\end{eqnarray}
where $N_{\rm Total}$ is the total number of events crossing this bar based on the direction provided by the STK, and $N_{\rm Signal}$ is the number of events that exhibit a signal (5$\sigma$ above the mean pedestal value).

\begin{figure}
\centering
\includegraphics[width=130mm,height=90mm]{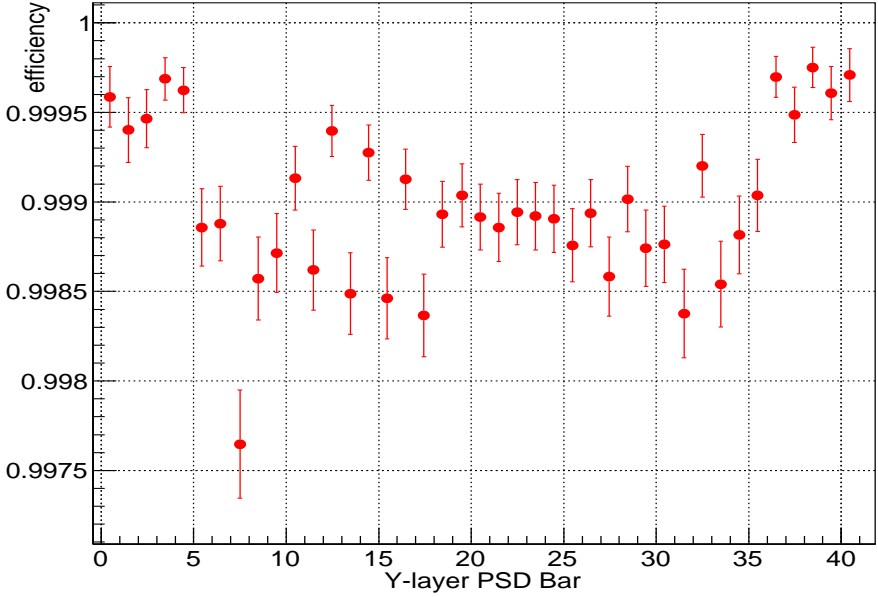}
\caption{Charged particle detection efficiency of the 41 Y-layer PSD bars obtained using proton MIPs collected during DAMPE flight operations.}
\label{Fig7}
\end{figure}

Fig.~\ref{Fig7} shows the efficiencies of all the 41 PSD bars in the $Y$ layer for events penetrating the entire PSD bar from top to bottom. We have measured that the detection efficiency is always better than $99.7\%$. In order to improve the background rejection, we combine the information of both PSD layers to form our charged particle veto as discussed in the following.

\begin{figure}
\centering
\includegraphics[width=130mm,height=90mm]{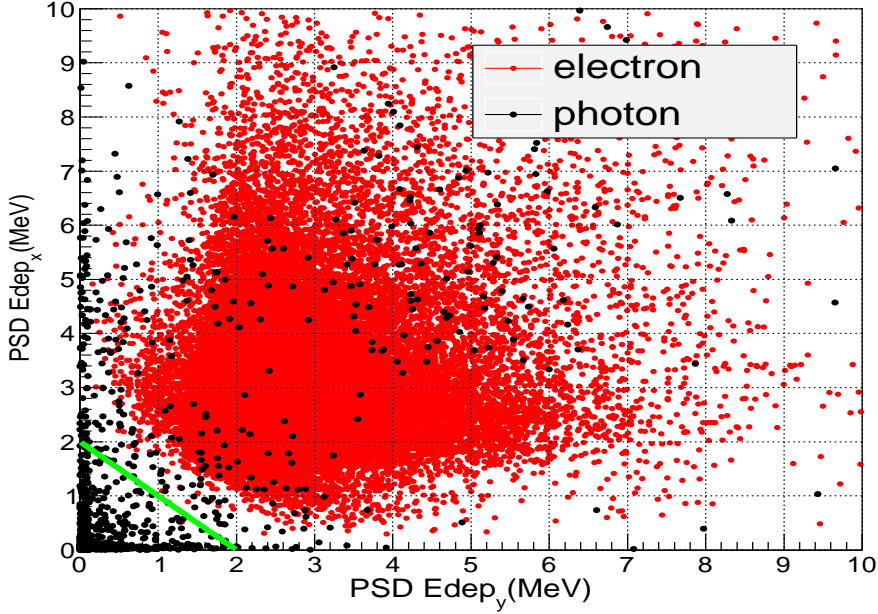}
\caption{Largest energy deposit in the selected PSD bars of the events in $Y$- and $X$-layer of PSD, for electrons (red) and $\gamma$-rays (black) in the [1 - 100]GeV energy range. The green line indicates the cut to select $\gamma$-rays.}
\label{Fig8}
\end{figure}

As previously discussed, the detection efficiency of $\gamma$-rays may be adversely affected by particle backsplash. We show in Fig.~\ref{Fig8} the largest deposit energy among the PSD bars of the X ($\rm Edep_x$) and Y ($\rm Edep_y$) layer, for MC electrons and $\gamma$-rays, respectively. The energy deposit due to backscattering of secondary particles for $\gamma$-rays is in general much smaller than that of electrons. Thus, we use the sum of $\rm Edep_x$ and $\rm Edep_y$ among the selected bars as the discriminating variable
to select $\gamma$-rays (shown by the green line in Fig.~\ref{Fig8} for an illustration).

\begin{figure}
\centering
\includegraphics[width=130mm,height=90mm]{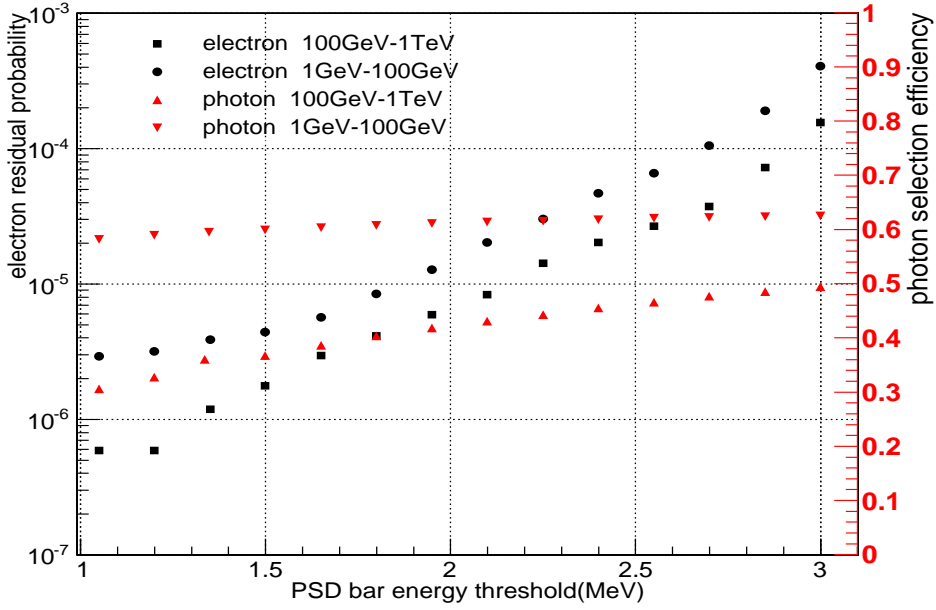}
\caption{Residual electron probability (left axis) and $\gamma$-ray detection efficiency (right axis) after the whole $\gamma$-ray selection procedure.}
\label{Fig9}
\end{figure}

The $\gamma$-ray detection efficiency of this whole algorithm is limited by the fact that a certain fraction ($\sim30\%$) of $\gamma$-rays penetrate STK without converting in $e^+e^-$ pairs. The detection efficiency also depends on the threshold used to define an energy deposit cluster for particles passing through the PSD. A higher threshold leads to a lower rejection of charged particles, whereas a low threshold potentially vetoes legitimate $\gamma$-ray events. To determine the optimal value of the threshold, we plot in Fig.~\ref{Fig9} the residual electron fraction (left axis) and $\gamma$-ray detection efficiency (right axis) as functions of the PSD threshold energy. Both the residual electron fraction and $\gamma$-ray efficiency increase with the threshold energy and decrease at higher energies because of the increasing amount of backscattering.
To balance the rejection power of electrons and the detection efficiency of $\gamma$-rays, we apply a threshold on $\rm Edep_x$+$\rm Edep_y$ of 1.5 MeV for GeV events, that yields an contamination less than $10^{-5}$ after all the selection cuts. Since the electron spectrum is softer than that of $\gamma$-rays at high energies (Baldini ~\cite{Baldini14}), we apply an energy-dependent threshold which linearly increases to 2.8 MeV at 1 TeV, maximizing the $\gamma$-ray detection efficiency keeping the electron contamination at a similar amount than that at low energies.

\section{Selection Acceptance}
\begin{figure}
\centering
\includegraphics[width=130mm,height=90mm]{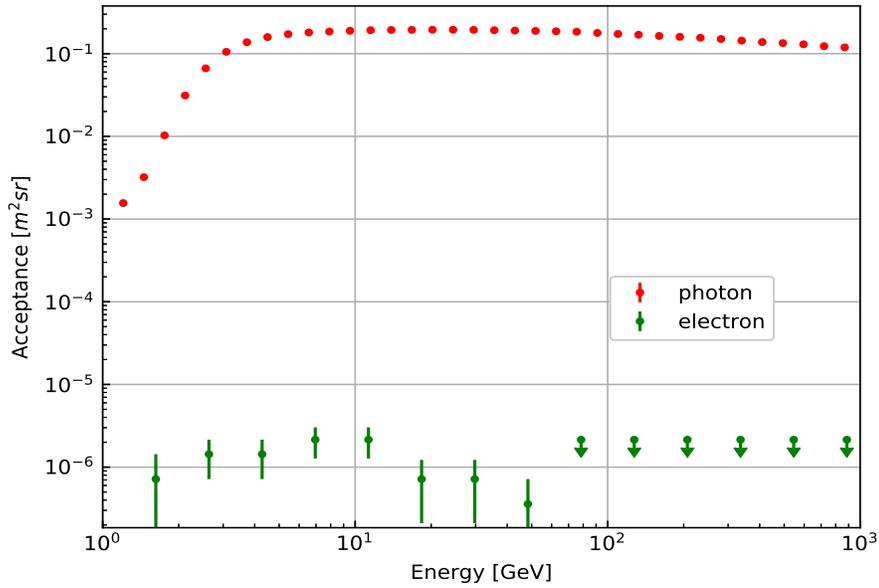}
\caption{Acceptance for $\gamma$-rays and for residual electrons after the $\gamma$-ray selection procedure developed in the context of this work.}
\label{Fig10}
\end{figure}

\begin{figure}
\centering
\includegraphics[width=130mm,height=90mm]{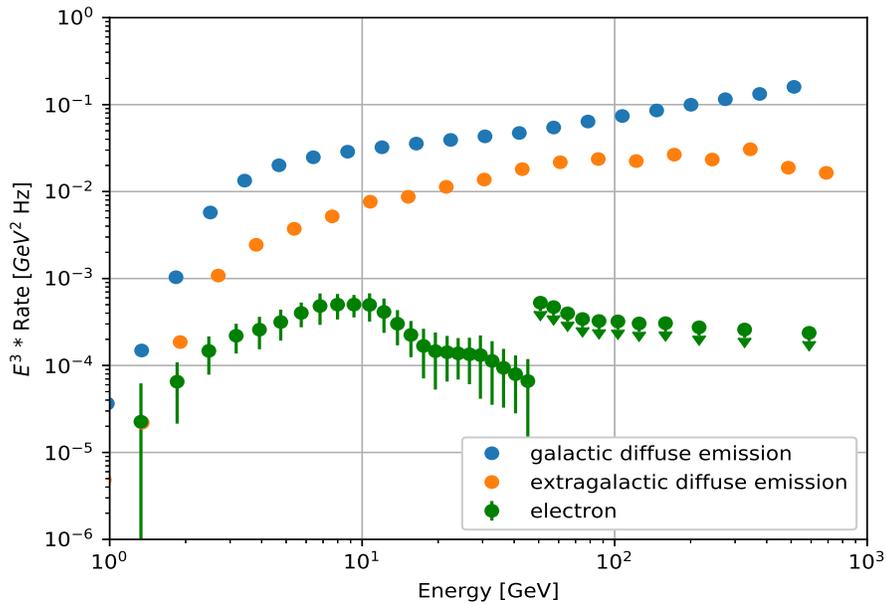}
\caption{Event rates of all-sky $\gamma$-rays and residual electrons predicted for DAMPE after applying the $\gamma$-ray selection developed in this work.}
\label{Fig11}
\end{figure}

We show in Fig.~\ref{Fig10} the acceptance of $\gamma$-rays and residual electrons after our selection procedure.
The effective acceptance of $\gamma$-ray observations is found to be $\sim0.19$ and $\sim0.11\,\mathrm{m^{2}sr}$ at 10 and $10^3\,\mathrm{GeV}$ respectively, and drops quickly at low energies due to the pre-scaling of low energy trigger (Chang et al. ~\cite{Chang17}).
In most of the energy range the acceptance of residual electrons is more than five orders of magnitude
smaller than that of $\gamma$-rays, as a consequence of an electron rejection factor of $\sim10^5$.
Convolving the acceptance with the electron energy spectra measured by AMS-02 (Aguilar et al. ~\cite{Aguilar14}) and the flux of the diffuse Galactic $\gamma$-ray emission (Acero et al. ~\cite{Acero16}) as well as the extragalactic $\gamma$-ray background (Ackermann et al. ~\cite{Ackermann15}), both measured by the Fermi-LAT, we obtain our predicted event rates as shown in Fig.~\ref{Fig11}. The electron background in the $\gamma$-ray selected sample is expected to be of the order of 1\% for energies higher than 5 GeV. Due to limited statistics of higher energy electron MC simulations $(E>50\,\mathrm{GeV})$, we cannot provide an accurate value of the residual electron background, thus electing to quote 95\% CL upper limits on these values instead.
The combination of the ACD and e/p separation capabilities of DAMPE results in a rejection factor better than $>10^7$ for protons.
We roughly estimate the contamination of protons and higher charge nuclei in the $\gamma$-ray sample to be significantly smaller compared to the potential electron contamination. In conclusion, we confirm that $\gamma$-ray identification algorithm outlined in this contribution results in a purity of the $\gamma$-ray sample better than 99\% in most of the energy range, providing a tool to select $\gamma$-rays for high accuracy measurements of $\gamma$-ray spectra and $\gamma$-ray astronomy.

\section{Verifying the $\gamma$-ray selection method using DAMPE flight data}
The instrument performance based on this $\gamma$-ray selection procedure has been already studied using simulations (Chang et al. ~\cite{Chang17}).
We discuss here the application of the $\gamma$-ray selection procedure to the DAMPE flight data to perform a data-driven verification of the $\gamma$-ray selection presented in the previous section.

\begin{figure*}[!ht]
  \centering
  \subfloat[]{
    \includegraphics[width=.5\textwidth,height=80mm]{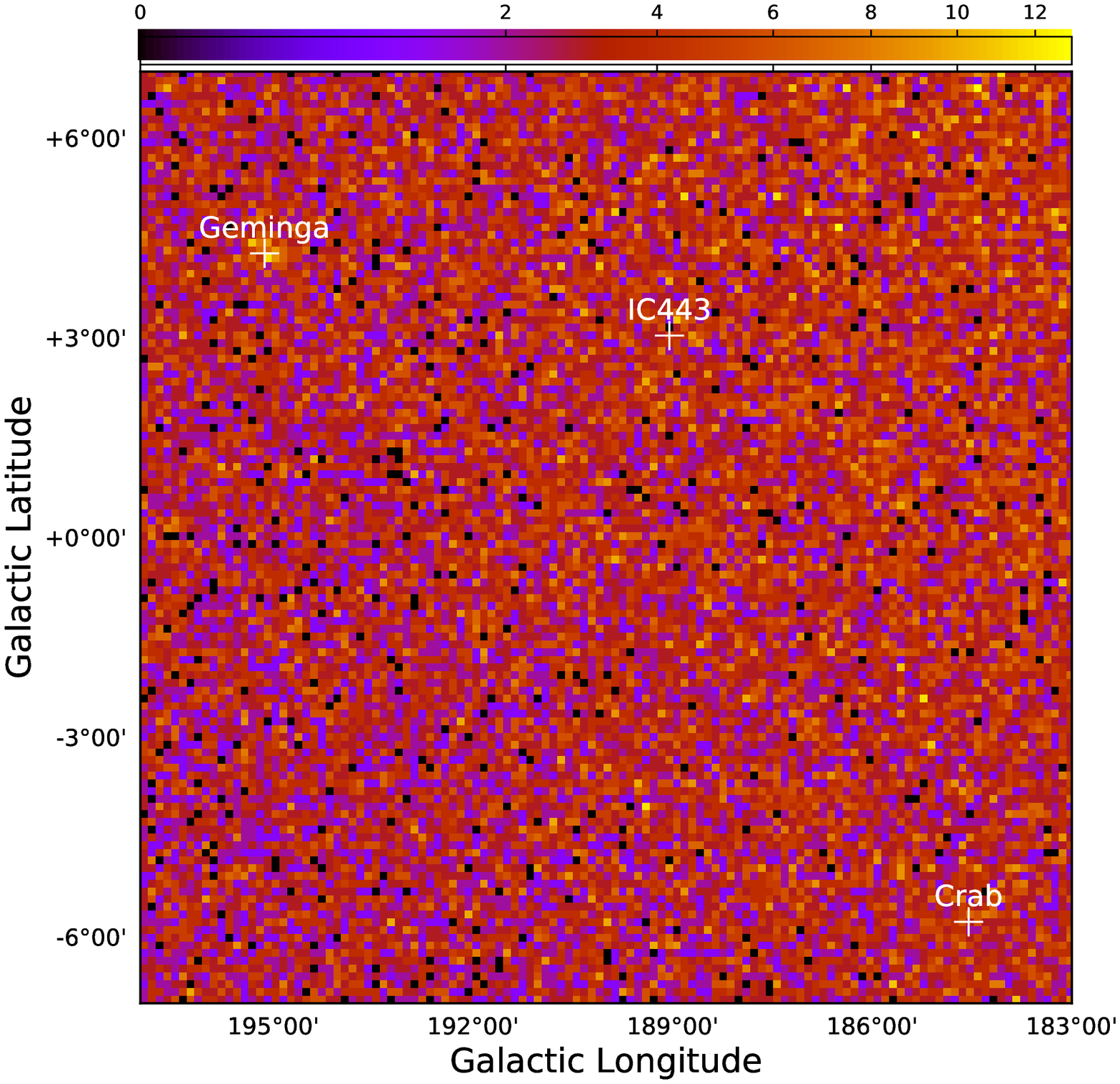}
  }
  \subfloat[]{
    \includegraphics[width=.5\textwidth,height=80mm]{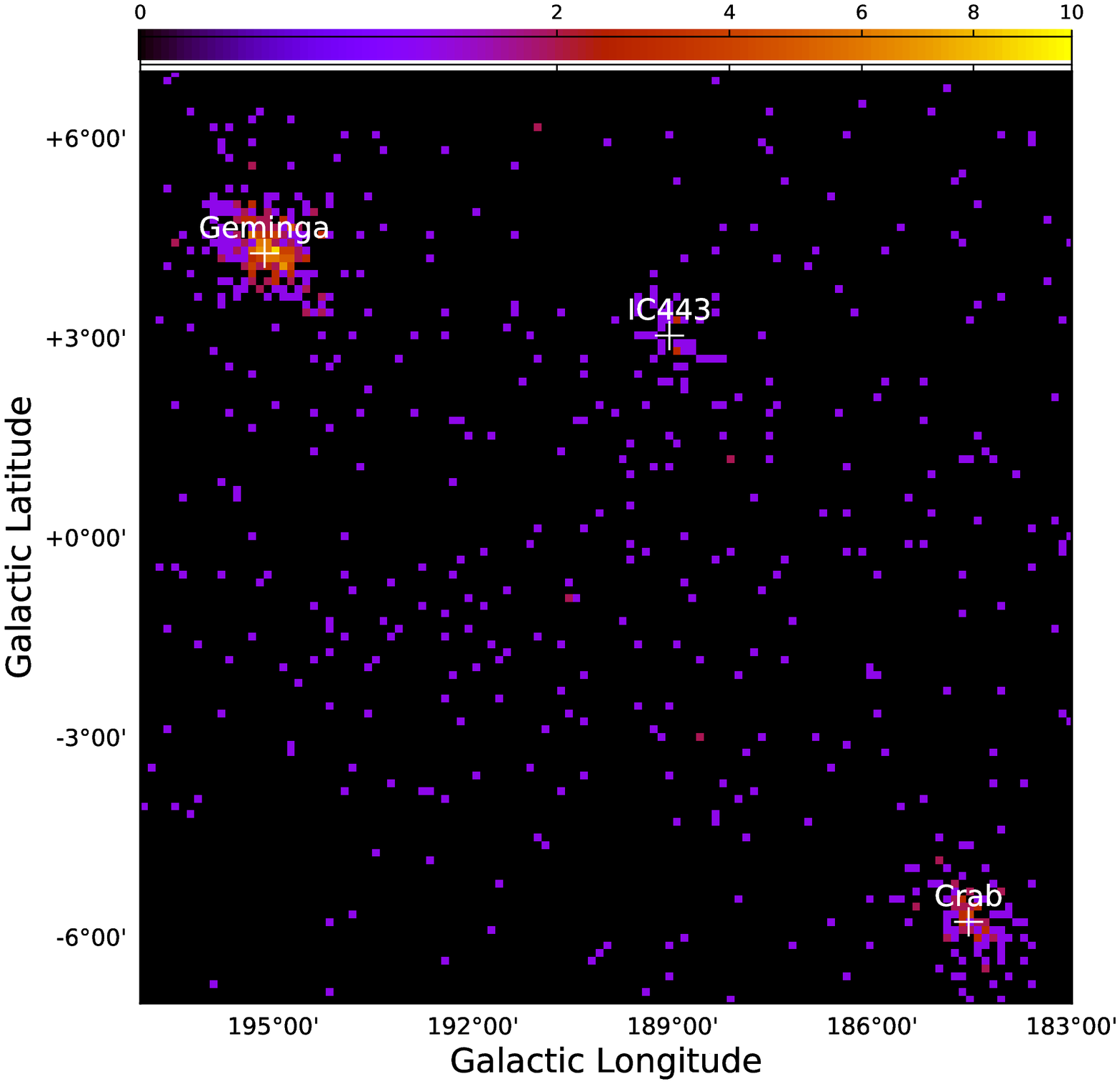}
  }
  \caption{(a) DAMPE sky map for a $15^{\circ}\times15^{\circ}$-region around Geminga, Crab and IC 443 built using all events collected in 1 year of data taking. (b) DAMPE sky map for the subset of $\gamma$-ray candidates for an energy range of 1 GeV to 1 TeV. The white crosses in the maps represent the coordinates of the sources as reported in SIMBAD.\protect\footnotemark}
  \label{Fig12}
\end{figure*}

DAMPE flight data collected from the period from 2016-01-01 to 2017-01-01 during which the detector has been operated in survey mode have been analyzed. $\gamma$-ray events in the energy range 1 GeV - 1 TeV have been selected excluding data taking periods in the South Atlantic Anomaly. Fig.~\ref{Fig12} shows the sky map in a $15^{\circ}\times15^{\circ}$ sky region containing several known bright sources.
The total count map, dominated by the nearly uniform background of charged CRs, is shown in the left panel.
The right panel shows the count map of $\gamma$-ray candidates, after the selection procedure. Three maxima, corresponding to the known $\gamma$-ray emitters Geminga, Crab, and IC 443 are clearly visible above the background.\footnote{We compare the maxima with the coordinates reported in SIMBAD (Wenger et al. ~\cite{Wenger2000}).} The sharp signatures in Fig.~\ref{Fig12} provide a first validation of the correctness of the $\gamma$-ray selection algorithm.

\footnotetext{http://simbad.u-strasbg.fr/simbad/}

\begin{figure}
\centering
\includegraphics[width=130mm,height=90mm]{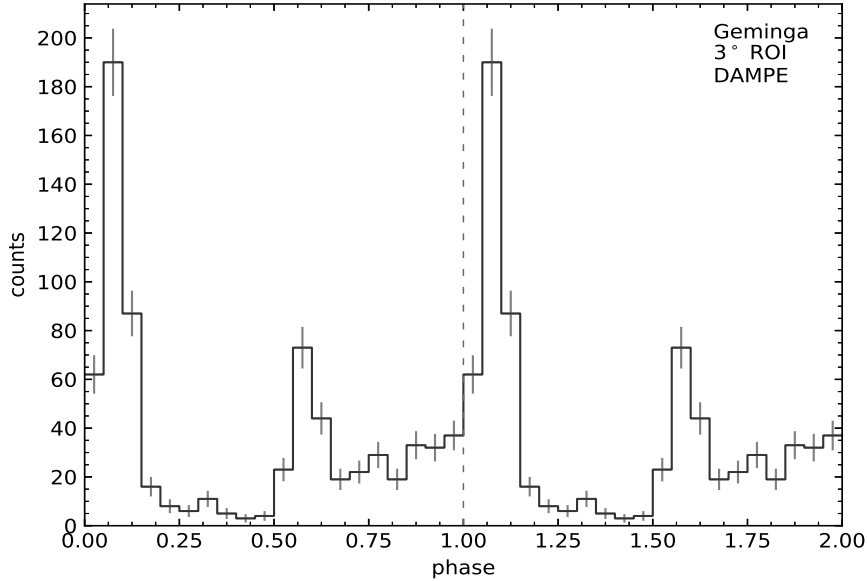}
\caption{Pulse profile of Geminga reconstructed using $\gamma$-rays candidates for an energy range of 1 GeV to 1 TeV identified in the DAMPE flight data . Error bars represent statistical uncertainties only.}
\label{Fig13}
\end{figure}

As one of the brightest $\gamma$-ray sources in the sky, we also perform an analysis of the Geminga pulsar, using 17 months of DAMPE data from 2016-01-01 to 2017-06-01. We select $\gamma$-ray candidate events for an energy range of 1 GeV to 1 TeV within $3^\circ$ around Geminga ($\alpha_{\rm 2000}=98.4756^\circ$, $\delta_{\rm 2000}=17.7703^\circ$; (Caraveo et al. ~\cite{Caraveo98}). The pulsar timing software TEMPO2 (Hobbs et al. ~\cite{Hobbs06}) is employed for the pulse-folding, based on the ephemeris obtained from the Fermi-LAT $\gamma$-ray data (Ray et al. ~\cite{Ray11}; Kerr et al. ~\cite{Kerr15}).\footnote{http://fermi.gsfc.nasa.gov/ssc/data/access/lat/ephems/. The spin frequency is increased by $4\times10^{-9}\ {\rm Hz}$ to improve the accuracy of the ephemeris, according to the Fermi-LAT data covering the period 2016-01-01 to 2017-06-01.} The resulting pulsar phase profile is shown in Fig.~\ref{Fig13}. We find that the phase profile extracted from DAMPE data is consistent with that reported by Fermi-LAT (Abdo et al.~\cite{Abdo2010}), where the relative difference observed in the intensity of the profile peaks is ascribed to the different energy spectra to which the two detectors are most sensitive. This comparison provides a complementary additional verification of the effectiveness of the $\gamma$-ray selection discussed in this contribution.

These preliminary results based on data-driven applications show that the procedure developed on MC simulations to identify $\gamma$-rays in DAMPE data is robust and effective. The selection algorithm will be adapted for the analysis of the future data that will be collected by DAMPE in the next years to achieve high-precision measurements of $\gamma$-ray sources and of the $\gamma$-ray sky.

\section{Summary}
Particle identification is one of the most important tasks of DAMPE data analysis. We have developed in this paper a method to identify converting $\gamma$-rays and separate them from charged CRs with DAMPE. A very efficient e/p separation algorithm, based on the different shower characteristics of electrons and $\gamma$-rays with respect to protons in the BGO calorimeter is first applied to identify electrons and $\gamma$-rays. After the trigger selection and the BGO e/p selection, 99.9\% of protons in the BGO fiducial volume are rejected.
We then identify the primary track associated to the incoming cosmic ray to minimize systematic effects due to the multiplicity in the number of tracks in the STK. Finally, the veto capability for charged particles of the PSD is employed to effectively suppress electrons and the remaining residual protons from the selected $\gamma$-ray samples. We estimate the rejection power and the detection efficiency (acceptance) based on detailed MC simulations of our detector. The effective acceptance of $\gamma$-ray observations is found to be $\sim0.19$ and $\sim0.11\,\mathrm{m^{2}sr}$ at 10 and $10^3\,\mathrm{GeV}$.
The rejection factors amount to $\sim10^{5}$ and $\sim10^{7}$ for electrons and protons, respectively.
By applying the event selection to DAMPE flight data, we can successfully identify well known bright $\gamma$-ray emitters and reconstruct the phase diagram for the Geminga pulsar in agreement with the report by Fermi-LAT, indicating a good consistency and consequently validating the event selection to be used for future high accuracy measurement of $\gamma$-rays physics with DAMPE.

\begin{acknowledgements}
The DAMPE mission was founded by the strategic priority science and technology projects in space science of the Chinese
Academy of Sciences (No. XDA04040000 and No. XDA04040400).
This work is supported in part by the National Key Research and Development Program of China (2016YFA0400200), the National Basic Research Program of China(No. 2013CB837000), the Strategic Priority Research Program of the Chinese Academy of Sciences ``Multi-Waveband Gravitational Wave Universe''(No. XDB23040000), the National Natural Science Foundation of China(Nos. 11525313, 11673075, 11773086, 11303107, 11303105, 11773085, U1738123, U1738136, U1738207 and U1738210), the Young Elite Scientists Sponsorship program by CAST No YESS20160196 and the 100 Talents Program of Chinese Academy of Sciences. In Europe DAMPE activities receive generous support by the Swiss National Science Foundation (SNSF), Switzerland and the National Institute for Nuclear Physics (INFN), Italy.

\end{acknowledgements}

\label{lastpage}


\begin{thebibliography}{99}
\bibitem[2010]{Abdo2010} Abdo A.A. et al (Fermi-LAT collaboration), 2010, Astrophy. J., 720, 1
\bibitem[2016]{Acero16} Acero F. et al. (Fermi-LAT collaboration), 2016, Astrophys. J. Suppl. Ser.,223,2
\bibitem[2015]{Ackermann15} Ackermann M. et al. (Fermi-LAT collaboration), 2015, Astrophy. J., 799, 1
\bibitem[2003]{Agostinelli03} Agostinelli S., Allison J., Amako K. et al., 2003, Nucl. Instrum. Methods A, 506, 250
\bibitem[2014]{Aguilar14} Aguilar M. et al. (AMS Collaboration), 2014, Phys. Rev. Lett., 113, 221102
\bibitem[2014]{Baldini14} Baldini L., 2014, arXiv:1407.7631
\bibitem[1998]{Barwick98} Barwick S. W., Beatty J. J., Bower C. R. et al., 1998, Astrophy. J., 498, 1998
\bibitem[1998]{Caraveo98} Caraveo P. A., Lattanzi M. G., Massone, G. et al,1998, Astron. Astrophys., 329, L1
\bibitem[2014]{Chang14} Chang J., 2014, Spac. Sci., 34, 550
\bibitem[2008]{Chang08} Chang J., AdamsJr J. H., Ahn H. S. et al., 2008, Adv. Spac. Res., 42, 431
\bibitem[2017]{Chang17} Chang J., Ambrosi G., An Q. et al., 2017, Astropart. Phys., 95, 6
\bibitem[2006]{Hobbs06} Hobbs G. B., Edwards R. T., Manchester R. N., 2006, Mon. Not. R. Astron. Soc., 369, 655
\bibitem[2015]{Kerr15} Kerr M., Ray P. S., Johnston S. et al., 2015, Astrophys. J., 814, 128
\bibitem[2011]{Ray11} Ray P. S., Kerr M., Parent D. et al., 2011, Astrophys. J. Suppl. Ser., 194, 17
\bibitem[2017]{Wang17} Wang C., Liu D., Wei Y.-F. et al., 2017, Chin. Phys. C., 41, 106201
\bibitem[2000]{Wenger2000} Wenger et al., 2000, Astron. Astrophys., 143, 9
\end{thebibliography}
\end{document}